\documentclass[11pt]{article}
\usepackage[margin=1in]{geometry}
\usepackage{amsmath,amssymb,amsthm}
\usepackage{booktabs}
\usepackage{hyperref}
\usepackage{xcolor}
\usepackage{listings}
\usepackage{microtype}
\usepackage{parskip}
\usepackage{titlesec}
\usepackage{enumitem}
\usepackage{cite}
\usepackage{graphicx}
\usepackage{caption}
\usepackage{array}
\usepackage{tabularx}
\usepackage{pgfplots}
\pgfplotsset{compat=1.18}
\usepackage{pgfplotstable}
\usepackage{float}
\usepackage{xurl}
\usepackage{tikz}
\usetikzlibrary{arrows.meta, positioning, calc}
\definecolor{purpleBox}{RGB}{75,58,168}
\definecolor{tealBox}{RGB}{15,90,78}
\definecolor{grayBox}{RGB}{74,74,74}
\definecolor{purpleSub}{RGB}{200,190,240}
\definecolor{tealSub}{RGB}{150,220,205}
\hypersetup{
 colorlinks=true,
 linkcolor=blue!60!black,
 citecolor=blue!60!black,
 urlcolor=blue!60!black
}
\title{\textbf{Structured Belief State and the First Precision-Aware Evaluation of LLM Memory Retrieval}}
\author{
 Jeffrey Flynt\\
 \textit{Independent Researcher}\\[0.3em]
 \href{mailto:jeffrey.flynt@utexas.edu}{\texttt{jeffrey.flynt@utexas.edu}}\\[0.2em]
}
\date{May 2026}
\begin{document}
\maketitle
\begin{abstract}
	
	Current LLM memory benchmarks evaluate answer quality rather than retrieval accuracy. Consequently, a system that dumps its entire belief store can achieve perfect recall and mask severe precision failures. We show this evaluation gap persists across multiple embedding models where similarity-based retrieval over domain-specific corpora inherently struggles to isolate target beliefs from semantically proximate ones. Furthermore, multi-turn topic drift compounds this retrieval noise while driving up latency and operational costs.
	
	To decouple retrieval quality from generative performance, we introduce PrecisionMemBench, an 89-case benchmark measuring precision, noise isolation, session latency, and belief mutability. We also present Tenure, a structured belief-state proxy that resolves scope and retrieval before inference and injects typed belief state as ambient instruction before the model sees the prompt, removing model-side discretion over whether memory is consulted. Evaluated across 13 configurations, Tenure passes all 77 single-turn cases and all 12 session cases. In contrast, the comparison configurations fail to reach even half of the active passes, with precision scores clustering at 0.22 and below. Our results demonstrate that while current memory systems successfully store information, they often fail to retrieve it cleanly, a structural vulnerability that traditional answer-quality benchmarks conceal.
\end{abstract}

\noindent\rule{\linewidth}{0.4pt}

Code: \href{https://github.com/tenurehq/tenure}{\texttt{github.com/tenurehq/tenure}} Benchmark: \href{https://github.com/tenurehq/precisionmembench}{\texttt{github.com/tenurehq/precisionmembench}}

\noindent\rule{\linewidth}{0.4pt}

\section{Introduction}
\label{sec:introduction}

LLM agent memory benchmarks, such as LongMemEval
\cite{wu2025longmemeval}, typically measure downstream answer quality
rather than upstream retrieval accuracy. A system can score
well by returning a broad candidate set and relying on a model to
identify relevant state during answer generation. Such evaluation can show that an answer remains recoverable, but does
not directly measure whether retrieval isolated the intended belief or
excluded state from another scope, an earlier lifecycle stage, or an
unrelated session turn.

Recent public disputes over memory benchmarks show that this evaluation
choice also complicates reproducibility. In a public discussion of
Mem0's~\cite{chhikara2025mem0} reported LongMemEval result, critics
identified benchmark-specific equivalence rules, hidden reasoning
instructions, and judge-side bias checks in the answer layer. Mem0's CTO
likewise argued that existing memory benchmarks place systems in
non-comparable harnesses and called for shared frameworks reporting cost
and latency alongside accuracy
~\cite{memory_benchmark_reproducibility_discussion_2026}. An independent
benchmark platform subsequently reported materially different LongMemEval
and LoCoMo~\cite{maharana2024locomo} results for Mem0 OSS under a public harness
~\cite{benchd_mem0}, with divergence traceable to differences in judge
model, token budget, and harness configuration rather than to
underlying retrieval architecture.

PrecisionMemBench scores returned belief IDs before answer
generation, separating retrieval behavior from downstream model and
judge behavior. It measures retrieval precision, noise isolation under
topic drift, session-turn latency, and same-session belief mutability.

Tenure addresses this problem by resolving identity, scope, and lifecycle
before inference rather than leaving those distinctions to the downstream
model.

\subsection{Identity Discrimination in Similarity Search}

Within a domain-specific belief store, semantic similarity may preserve
broad subject-matter relevance without uniquely identifying the intended
belief. Larger embedding models may alter
the ranking of those beliefs without adding explicit signals for identity,
scope, or supersession. This motivates separating candidate eligibility
from selection within the eligible set. In Tenure, scope constrains which
beliefs may be considered, while identity-oriented retrieval selects among
those candidates.

\subsection{Contributions}

\begin{enumerate}
	\item \textbf{A precision-aware retrieval benchmark}
	89 cases covering alias resolution,
	scope disambiguation, supersession, fuzzy matching, cross-user
	isolation, budget limits, ranking stability, and session-level noise
	isolation under topic drift. Each case specifies both required and
	prohibited results.
	
	\item \textbf{A structured belief representation}
	Belief Architecture with explicit type, scope,
	epistemic status, provenance, lifecycle state, and a
	\texttt{why\_it\_matters} field that preserves how stored facts should
	shape future responses.
	
	\item \textbf{A precision-first retrieval design}
    Retrieval grounded in named-entity resolution, using alias-weighted matching as an identity-oriented alternative to semantic similarity once vocabulary has converged.
	
	\item \textbf{A compaction architecture} Alias enrichment and supersession limit noise accumulation while
	preserving historical state.
\end{enumerate}

\section{Related Work}
\label{sec:related}

Recent memory systems add structure through fact extraction, typing, or
organization, but often retain similarity-based retrieval at read time.
This evaluation tests whether retrieval preserves distinctions introduced
during extraction. Mem0~\cite{chhikara2025mem0}, for example, extracts
natural-language facts and retrieves them through embedding similarity.

Memori~\cite{borro2026memori} also extracts structured state rather than
injecting raw transcripts. It represents dialogue as semantic triples
linked to conversation summaries and retrieves through cosine similarity
and BM25. Memori's triples record facts, whereas Tenure's beliefs also
specify how those facts should shape future responses.

A-MEM \cite{xu2025amem} treats memory as a self-organizing network of Zettelkasten-inspired notes where new memories trigger updates to the contextual representations of existing ones. The published architecture does not describe a supersession chain or audit log, nor an equivalent structural mechanism for retiring outdated context.

\subsection{Structured Memory Evaluation}

StructMemEval~\cite{shutova2026structmemeval} evaluates an agent's ability
to organize long-term memory rather than merely retrieve facts, isolating
tasks requiring state tracking, hierarchical organization, and accumulated
counting. Its central claim is that retrieval-augmented systems fail these tasks
even when the underlying facts are available and the retrieval budget
is generous. Like LoCoMo~\cite{maharana2024locomo}, StructMemEval
assesses correctness at the answer layer rather than directly scoring
the retrieved set.

\subsection{Dual-Route Retrieval and Structural Discrimination}

H-Mem~\cite{ye2026hmem} and Mnemis~\cite{tang2026mnemis} independently
supplement similarity-based retrieval with structural retrieval mechanisms. H-Mem uses parallel temporal and semantic trees, while Mnemis combines
System-1 similarity search with System-2 Global Selection over a
hierarchical graph. The pattern
across recent high-performing memory systems is convergent, with more
retrieval paths, heavier embedding infrastructure, and re-ranking
stages layered on top of similarity search.
Tenure's position is that the additional infrastructure is compensating for
the wrong primary signal.

\subsection{Named Entity Resolution and BM25}

Our choice of BM25 over embedding similarity is motivated by work on
named entity resolution
\cite{robertson1994some,mihalcea2007wikify}. Within a domain, the terms
used to identify an entity may provide a more discriminative retrieval
signal than broad semantic relatedness alone. Embedding similarity
emphasizes semantic relatedness, whereas our alias-weighted BM25
configuration emphasizes terms associated with entity identity within
the domain.

\section{The Belief Architecture}
\label{sec:belief-architecture}

A \textit{belief} is the atomic unit of persistent context. The term is a deliberate borrowing from the Belief layer of the Belief-Desire-Intention (BDI) agent architecture~\cite{rao1995bdi}, which treats beliefs as an agent's informational state, distinct from its goals and action plans.

Tenure is deployed as an inference proxy rather than a model-invoked
memory tool. Before inference, the proxy resolves scope, retrieves
eligible beliefs, injects structured state into the model context, and
forwards the request to the provider. This removes model-side discretion
over whether memory is consulted.

Hindsight~\cite{latimer2025hindsight} reported a similar control-flow
problem where a model-selected MCP integration sometimes answered directly or
completed turns without invoking memory, leading the integration to adopt
automatic pre-inference recall and post-turn retention. This experience
supports placing memory that must reliably shape inference in the
execution path around the model rather than behind an optional tool
call.

\begin{figure}[t]
	\centering
	\begin{tikzpicture}[
		font=\small\sffamily,
		box/.style={
			rectangle,
			rounded corners=2pt,
			draw=none,
			text width=1.48cm,
			minimum width=1.72cm,
			minimum height=1.25cm,
			align=center,
			inner xsep=1mm,
			inner ysep=1.5mm
		},
		gray/.style={box, fill=grayBox, text=white},
		purple/.style={box, fill=purpleBox, text=white},
		teal/.style={box, fill=tealBox, text=white},
		lbl/.style={
			font=\small\sffamily\bfseries,
			text=black,
			anchor=west
		},
		arr/.style={
			-{Latex[length=1.4mm, width=1mm]},
			semithick,
			gray!70!black
		},
		edgelbl/.style={
			font=\small\sffamily\bfseries,
			text=black!70,
			align=center,
			fill=white,
			inner sep=0.5pt
		}
		]
		\node[lbl] (t1) at (0,0) {Typical memory (optional model-gated)};
		
		\node[gray, anchor=north west] (q1) at (0,-0.35) {
			User query
		};
		\node[purple, right=3.5mm of q1] (m1) {
			\textbf{Model}
		};
		\node[purple, right=3.5mm of m1] (s1) {
			\textbf{Memory MCP tool}
		};
		\node[gray, right=3.5mm of s1] (r1) {
			Response
		};
		
		\draw[arr] (q1) -- (m1);
		\draw[arr] (m1) -- (s1);
		\draw[arr] (s1) -- (r1);
		
		\node[lbl, anchor=west] (t2) at (0,-2.25) {
			Tenure (always injected)
		};
		
		\node[gray, anchor=north west] (q2) at (0,-2.60) {
			User query
		};
		\node[teal, right=3.5mm of q2] (m2) {
			\textbf{Tenure}\\[1pt]
		};
		\node[teal, right=3.5mm of m2] (s2) {
			\textbf{Model}\\[1pt]
		};
		\node[gray, right=3.5mm of s2] (r2) {
			Response
		};
		
		\draw[arr] (q2) -- (m2);
		\draw[arr] (m2) -- (s2);
		\draw[arr] (s2) -- (r2);
	\end{tikzpicture}
	\caption{Request paths for model-gated tool retrieval (top) and Tenure's always-injected proxy (bottom).}
	\label{fig:request-path}
\end{figure}
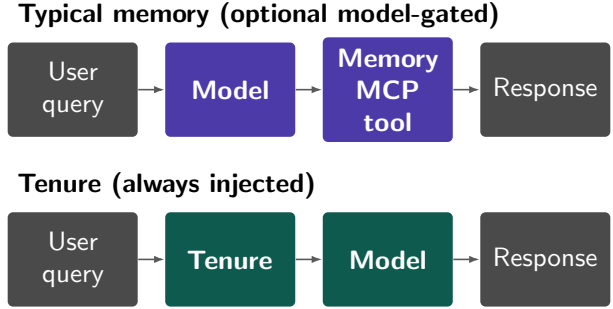

Figure~\ref{fig:request-path} contrasts the two request paths.

Tenure stores extracted conclusions as typed, scoped, and versioned
beliefs containing epistemic status, confidence, provenance, canonical
name, aliases, lifecycle state, and supersession links.

\subsection{The \texttt{why\_it\_matters} Field as Pre-Computed Action}

Every belief includes a \texttt{why\_it\_matters} field specifying how
the conclusion should shape future responses. If extraction cannot
produce that instruction, no belief is created. For example, rather than
recording only ``Uses TypeScript with strict mode,'' the belief instructs
future responses to use strict TypeScript without implicit
\texttt{any}.

\subsection{Persona Prelude}

The persona prelude is a compact, cached projection over active and
inferred preference beliefs in the universal user scope. It summarizes
expertise calibration, communication style, and working style rather
than injecting the contributing records individually. Exploratory
beliefs are excluded from this projection.

At inference time, the prelude follows a separate route from pinned
facts, query-dependent beliefs, and open questions. Stable interaction
guidance remains available across topics without competing
with project-specific beliefs for retrieval slots.

\subsection{Epistemic State, Lifecycle, and Scope}

Epistemic status distinguishes stated, inferred, and exploratory beliefs.
Lifecycle and scope independently control retrieval eligibility. Resolved
and superseded records are excluded before textual ranking, while domain
and project beliefs are eligible only when their scope is active.
Superseded beliefs remain available for historical inspection. Stable
universal preferences follow the persona route rather than competing in
query-dependent retrieval.

\section{Retrieval Design}
\label{sec:retrieval}

A belief whose canonical
name is \texttt{kubernetes} and whose aliases include \texttt{k8s} and
\texttt{kube} should be retrievable from a query containing
\texttt{k8s} because that surface form is already part of the belief.

Alias-weighted BM25 is weaker on terms the system has not seen before.
This design favors an initial miss over returning nearby but irrelevant
beliefs. A coworker who has never heard a term usually asks
what it refers to rather than guessing based on resemblance to
something else already in the room. The first miss serves a similar
role here. Once the new wording is captured as an alias, later queries
can resolve through it directly. The \textit{Iterative Alias Enrichment} section describes that alias enrichment process.

\subsection{Search Architecture and Index Engineering}

Chat retrieval inverts the usual BM25 setting: queries may contain
hundreds of tokens, while canonical names and alias lists are short.
Tenure decomposes structured canonical names into matchable
tokens, preserves alias surfaces without stemming, indexes multi-word
aliases as phrases, and boosts canonical and alias-phrase matches. This
allows a precise identity term to outweigh repeated terms from a long
natural-language query.

\paragraph{Retrieval Pipeline}

Retrieval first filters by ownership, active scope, and lifecycle state,
then applies alias-weighted ranking within the eligible set. Persona
guidance, pinned beliefs, query-dependent beliefs, and unresolved
questions follow separate routes. Identified relation participants are
loaded structurally rather than recovered through another similarity
search. The resulting state is allocated under the context budget and
injected as typed sections before inference.

\section{Conflict Resolution and Compaction}
\label{sec:compaction}

A new extraction may create, reinforce, or conflict with a belief.
Compaction merges duplicate state and combines aliases, preserving the
vocabulary learned through prior interactions. When one belief replaces
another, the earlier belief remains linked in the historical record but
is excluded from future injection. Outdated terminology can still resolve
through the supersession chain to the active belief.

\section{Evaluation}
\label{sec:evaluation}

The benchmark contains 77 single-turn cases and 12 turn-level assertions across two session scenarios. Each case specifies a requesting user, one or more active scopes, a natural-language query, an optional belief budget, and the belief IDs that must be returned or excluded.

Expectations use positive and negative assertions. \texttt{mustInclude} requires an ID to appear, \texttt{mustExclude} prohibits an ID, and \texttt{shouldOnlyInclude} defines the complete allowable set. A case passes only when all asserted ID constraints are satisfied. Retrieving the target does not produce a pass if the result also contains a prohibited belief or omits another required belief.

Retrieval precision and recall are computed over returned query-dependent belief IDs only when a case requires active retrieval.

\paragraph{Pass Taxonomy.}
The 77 single-turn cases distinguish three forms of passing behavior.
An \textit{active retrieval pass} requires the complete allowable set
of query-dependent beliefs. A \textit{structural pass} verifies scope,
supersession, type, or related constraints without necessarily requiring
query-dependent retrieval. A \textit{trivially empty pass} expects no
query-dependent beliefs because of the query or available budget. The
suite contains 43 active, 25 structural, and 9 trivially empty cases.

We evaluate Tenure against 12 comparison configurations: the Open
Knowledge Format reference agent~\cite{okf_github},
Gbrain~\cite{gbrain_github},
Supermemory~\cite{supermemory_github},
YourMemory~\cite{yourmemory_github},
Cognee~\cite{markovic2025optimizinginterfaceknowledgegraphs},
A-MEM~\cite{xu2025amem},
AtomicMemory~\cite{atomicmemory_github},
Hindsight~\cite{latimer2025hindsight},
Mem0~\cite{chhikara2025mem0},
Zep~\cite{rasmussen2025zep},
AgentMemory~\cite{agentmemory_github}, and an in-harness vector baseline.
Where the harness controlled embedding choice, embedding-based runs used
\textbf{mxbai-embed-large}~\cite{emb2024mxbai,li2023angle} at 1024
dimensions.

\begin{table}[t]
	\centering
	\small
	\setlength{\tabcolsep}{3.5pt}
	\begin{tabular}{lrrr}
		\toprule
		\textbf{Provider} &
		\textbf{Active} &
		\textbf{Structural} &
		\textbf{Empty} \\
		\midrule
		Tenure            & 43 & 25 & 9 \\
		Open Knowledge Format & 18 & 13 & 5 \\
		Gbrain                &  5 & 20 & 9 \\
		Supermemory           &  4 & 14 & 3 \\
		Yourmemory            &  0 & 15 & 6 \\
		Vector                &  0 &  8 & 3 \\
		Cognee                &  0 &  7 & 4 \\
		A-MEM                 &  0 &  6 & 3 \\
		Atomicmemory          &  0 &  6 & 3 \\
		Hindsight             &  0 &  6 & 3 \\
		Mem0                  &  0 &  6 & 3 \\
		Zep                   &  0 &  6 & 3 \\
		agentmemory           &  0 &  5 & 2 \\
		\bottomrule
	\end{tabular}
	\caption{Single-turn passes by assertion type. Only active passes demonstrate successful query-dependent retrieval. Structural passes verify filtering or routing behavior, while empty passes require no query-dependent result.}
	\label{tab:pass-taxonomy}
\end{table}

The benchmark makes retrieval correctness
testable in a field that lacks a settled definition of what AI memory should
return and exclude.

\paragraph{Seed Corpus as a Contrast Set.}
The 35-belief seed contains 23 code-scoped, 7 writing-scoped, and
5 universal beliefs, together with a secondary-user fixture. Controlled
collisions test overlapping names, scope, lifecycle state, supersession,
relations, and cross-user isolation. The benchmark scores required and
prohibited belief IDs without prescribing a retrieval architecture.

Table~\ref{tab:benchmark-categories} summarizes the benchmark categories
and the practical retrieval requirements they test.

\begin{table}[t]
	\centering
	\footnotesize
	\setlength{\tabcolsep}{5pt}
	\renewcommand{\arraystretch}{1.08}
	\begin{tabular}{
			@{}
			p{0.22\textwidth}
			p{0.72\textwidth}
			@{}
		}
		\toprule
		\textbf{Case category} &
		\textbf{Practical requirement tested} \\
		\midrule
		
		Alias resolution &
		Whether shorthand, abbreviations, and natural variations resolve to the intended belief without unrelated results. \\
		
		Scope disambiguation &
		Whether identical names remain isolated across projects and domains. \\
		
		Supersession chain exclusion &
		Whether changed decisions and preferences stop influencing current behavior while the historical record remains available for audit. Queries using outdated terminology must not restore obsolete guidance. \\
		
		Fuzzy matching and prefix guards &
		Whether ordinary misspellings and transpositions still reach the intended memory without allowing similarly spelled but unrelated terms to produce false matches. \\
		
		Counter-signal retrieval &
		Whether rejected or replaced terminology surfaces the current governing decision. \\
		
		Relation expansion &
		Whether relationship queries return eligible associated state without unrelated or out-of-scope participants. \\
		
		Session-level noise isolation &
		Whether returning to an earlier topic excludes memories introduced during intervening topic drift. \\
		
		Budget eviction and capacity &
		Whether limited context capacity is allocated to required state under competing retrieval demands. \\
		
		Design boundary cases &
		Whether empty, ambiguous, or unsupported queries produce predictable results without low-confidence retrieval. \\
		
		Type routing and open questions &
		Whether open questions remain distinct from established state and disappear from the active route when resolved. \\
		
		Ranking stability &
		Whether specific queries rank the intended belief ahead of weaker matches. \\
		
		Cross-user isolation &
		Whether one user's stored information remains inaccessible to another. \\
		
		Cold-start behavior &
		Whether a user with no stored state receives a clean empty result. \\
		
		Persona prelude content &
		Whether durable preferences remain available across topics without consuming query-dependent retrieval capacity. \\
		
		\bottomrule
	\end{tabular}
	\caption{Benchmark categories and practical retrieval requirements. Providers are scored on required and prohibited belief IDs without being required to adopt Tenure's architecture.}
	\label{tab:benchmark-categories}
\end{table}

Comparison systems use documented out-of-the-box configurations unless
a change is required by the common interface or a model dependency.
Providers requiring an LLM use Claude Sonnet 4.6. Each system is pinned
to a public revision, binary, or container digest, and wrappers may only
adapt the common \texttt{/add}, \texttt{/search}, and \texttt{/reset}
contract. Configurations that cannot be pinned or that alter provider
ranking, thresholds, fixtures, or retrieval behavior are excluded.

All represented providers were invited to verify their integrations,
request reruns, propose documented configurations, or submit fixes.
Accepted changes were rerun and incorporated into the reported results. In one case, an overnight iteration with YourMemory~\cite{yourmemory_github}, led to
multiple pull requests addressing diagnosed failure modes; we reran the
benchmark after the changes and incorporated the improved result into the
reported tables.

\paragraph{Latency Interpretation.}
Latency is measured at the benchmark interface each provider exposes.
For ingestion, this is the observed time for each \texttt{/add} call during
seeding. For retrieval, latency is the observed time for the provider's
\texttt{/search} call during evaluation. When an API does not expose stable belief IDs, the benchmark subsequently performs an untimed wrapper-side lookup to reconcile returned content with seeded IDs for scoring.

\paragraph{Common Evaluation Harness.}

The evaluation harness maps each provider's returned content to seeded
belief IDs when the provider does not expose stable IDs directly. This
normalization does not alter provider ranking, thresholds, or candidate
selection. Cases are scored against the resulting IDs, so comparison
systems are not required to implement Tenure's schema or
output structure.

\subsection{Retrieval Results}
\label{sec:retrieval-results}

Among comparison systems, the Open Knowledge Format records the highest mean
precision at 0.47 and the most active retrieval passes at 18.
Supermemory follows at 0.22 precision with 4 active passes. Most other
systems achieve substantially higher recall than precision, indicating
that targets are often returned alongside beliefs outside the allowable
set.

\subsubsection*{Retrieval Precision on Active-Assertion Cases}

Table~\ref{tab:precision-active-expanded} reports mean retrieval
precision and recall restricted to cases where
\texttt{retrievalPrecision} is not \texttt{null}.

\begin{table}[t]
	\centering
	\begin{tabular}{lcc}
		\toprule
		\textbf{Provider} &
		\textbf{Mean precision} &
		\textbf{Mean recall} \\
		\midrule
		Tenure            & 1.00 & 1.00 \\
		Open Knowledge Format & 0.47 & 0.91 \\
		Supermemory           & 0.22 & 0.71 \\
		agentmemory           & 0.17 & 0.97 \\
		Yourmemory            & 0.17 & 0.88 \\
		Atomicmemory          & 0.15 & 0.95 \\
		Gbrain                & 0.14 & 0.17 \\
		Zep                   & 0.09 & 0.95 \\
		Vector                & 0.09 & 1.00 \\
		Hindsight             & 0.06 & 1.00 \\
		Mem0                  & 0.06 & 0.99 \\
		A-MEM                 & 0.06 & 0.99 \\
		Cognee                & 0.05 & 0.92 \\
		\bottomrule
	\end{tabular}
	\caption{
		Mean retrieval precision and recall on active-assertion cases only.
	}
	\label{tab:precision-active-expanded}
\end{table}

\subsubsection*{Semantic Proximity and Belief Identity}

We tested cosine retrieval with three embedding models:
\texttt{nomic-embed-text}~\cite{nussbaum2024nomic} at 768 dimensions,
\texttt{mxbai-embed-large}~\cite{emb2024mxbai,li2023angle} at 1024
dimensions, and \texttt{qwen3-8b}~\cite{qwen3technicalreport} at 4096
dimensions. All three produce the same precision and pass count, while
\texttt{qwen3-8b} exceeds 1{,}100ms mean latency. Under this fixed cosine
policy, model scale does not resolve the observed identity failures.

\begin{table}[H]
	\centering
	\small
	\setlength{\tabcolsep}{3pt}
	\begin{tabular}{lcccrr}
		\toprule
		\textbf{Model} &
		\textbf{Prec.} &
		\textbf{Rec.} &
		\textbf{Passes} &
		\textbf{Mean} &
		\textbf{p95} \\
		\midrule
		nomic-embed-text  & 0.09 & 1.0 & 11/77 &    43.36 &    85.21 \\
		mxbai-embed-large & 0.09 & 1.0 & 11/77 &    96.48 &    257.24 \\
		qwen3-8b          & 0.09 & 1.0 & 11/77 &  1130.95 & 2604.84 \\
		\bottomrule
	\end{tabular}
	\caption{
		Vector retrieval across three embedding models. Latency is reported
		in milliseconds; none of the passes is an active retrieval pass.
	}
	\label{tab:embedding-invariance}
\end{table}

A related pattern appears in the evaluated Hindsight configuration. On
the alias-resolution and scope-disambiguation cases described below,
Hindsight~\cite{latimer2025hindsight} returns 18 beliefs while omitting
the target, producing precision 0 and recall 0. Although Hindsight's full
image includes a cross-encoder reranker intended to improve
discrimination, the evaluated configuration did not isolate the target
beliefs in these cases.

\subsubsection*{Structural Boundaries as Candidate Constraints}

When scope and user identity are not enforced as hard retrieval boundaries, a missed filter
path can let one user's or agent's memory surface in another's
context. AgentMemory
fails both scope disambiguation and cross-user isolation in the benchmark;
a later issue report identified omitted agent-scope filtering in its
shared \texttt{mem::search} path
\cite{agentmemory_github,agentmemory_issue_817}.

\subsubsection*{Divergence Between Storage and Retrieval}

Inspection of Mem0's stored representation showed that extraction
preserved a relation between an authentication service and Redis,
including Redis's role in session storage.

For the query \textit{``what are the auth service dependencies and failure
	modes?''}, the ground-truth set contains the relation and its Redis
participant. Mem0~\cite{chhikara2025mem0} retrieves the relation but not the
participant, while returning 17 unexpected records, yielding precision
$1/18 = 0.056$ and recall $1/2 = 0.5$. A more specific query retrieves
both records but still returns 16 unrelated beliefs.

When a stored relation already contains participant identifiers, its
eligible participants can be loaded structurally rather than recovered
through another semantic search.

\subsection{Retrieval Noise Under Topic Drift}
\label{sec:session-eval}

The 12 session cases test whether off-topic turns contaminate retrieval
when a session returns to its original topic. The main scenario opens on
one topic, drifts across eight unrelated turns, and returns at turns 9
and 10 using newly introduced surface forms. The \textit{drift score} is
the fraction of retrieved non-pinned beliefs originating in drift turns;
0 indicates perfect isolation.

\begin{table}[H]
	\centering
	\small
	\setlength{\tabcolsep}{2.5pt}
	\begin{tabular}{lrrrr}
		\toprule
		\textbf{Provider} &
		\textbf{Passed} &
		\textbf{Drift} &
		\textbf{Prec.} &
		\textbf{ms} \\
		\midrule
		Tenure    & 12/12 & 0.000 & 1.000 &   47.8 \\
		OKF           &  2/12 & 0.215 & 0.569 & 3349.5 \\
		Yourmemory    &  1/12 & 0.737 & 0.197 &  430.5 \\
		Supermemory   &  1/12 & 0.749 & 0.183 &  172.3 \\
		Cognee        &  1/12 & 0.846 & 0.077 & 4222.6 \\
		Gbrain        &  1/12 & 0.000 &   --- &  535.6 \\
		agentmemory   &  0/12 & 0.809 & 0.191 &   98.5 \\
		Atomicmemory  &  0/12 & 0.845 & 0.155 &  355.1 \\
		Zep           &  0/12 & 0.889 & 0.111 &  418.1 \\
		Vector        &  0/12 & 0.914 & 0.086 &  256.8 \\
		A-MEM         &  0/12 & 0.926 & 0.074 &   25.7 \\
		Hindsight     &  0/12 & 0.929 & 0.072 & 1880.6 \\
		Mem0          &  0/12 & 0.940 & 0.060 &  377.9 \\
		\bottomrule
	\end{tabular}
	\caption{Session results across 12 drift-sensitive cases. Drift is the
		fraction of retrieved non-pinned beliefs originating in off-topic turns;
		0 is perfect isolation. Precision is undefined when no relevant belief is
		retrieved.}
	\label{tab:session-expanded}
\end{table}

At turn 10, the vector baseline ranks \texttt{b-redis-code} first but
also retrieves all seven drift beliefs. Zep~\cite{rasmussen2025zep}
retains the target with 0.92 drift, showing that recall can coexist with
severe contamination. Hindsight~\cite{latimer2025hindsight} omits the
target and triggers four noise violations.

Hindsight averages
672.15ms for single-turn retrieval and 173.3 seconds to ingest the
35-belief corpus. Zep averages 139.64ms for retrieval but requires
897.0 seconds for ingestion, with individual beliefs taking up to
125{,}148ms to become available. Availability delay is reported
separately from retrieval precision.

\subsection{External Adoption and Diagnostic Use}
\label{sec:external-adoption}

Externally developed memory projects have used PrecisionMemBench to
report retrieval boundaries, diagnose failure modes, evaluate
implementation changes, and design retrieval-level test suites. 

ctrl-memory\cite{ctrl_memory_github} foregrounds PrecisionMemBench in its public benchmark
reporting and presents outcomes by capability rather than only as an
aggregate score. It improved from 9 passes with keyword retrieval to 42
with hybrid retrieval, 43 with fuzzy matching, 49 with scope filtering,
and 54 with supersession filtering. Its published analysis identifies
both supported behaviors and unresolved categories, including token
bleeding, relation expansion, capacity stress, ranking weights,
multi-scope collisions, fuzzy edge cases, and type routing. The project used these categories to guide
the incremental addition of retrieval, scope, and lifecycle controls.

Gbrain\cite{gbrain_github} applied the benchmark to a recall-oriented hybrid retriever and
identified broad return sets as a precision failure that recall alone did
not expose. The finding motivated an intent-aware return-sizing feature.
Further instrumentation showed that gaps between adjacent rank scores did
not reliably distinguish correct from incorrect top results, leading the
project to reject a proposed score-cliff mechanism. The benchmark therefore supported both an
implementation change and a negative design decision.

YourMemory\cite{yourmemory_github} incorporated PrecisionMemBench into its own repository through
a native runner and committed result artifacts. It used the benchmark to
investigate a discrepancy between its native execution path and an
earlier adapter path, localizing the disagreement to integration and
state management. This demonstrates that the
benchmark assertions can support provider-side reproduction and diagnosis
at the boundary between a memory system and its evaluation wrapper.

A fourth project, harness-mem\cite{harness_mem_github}, selected retrieval-isolated evaluation as
its highest-priority benchmark task after reviewing current agent-memory
and retrieval research. Its proposed golden suite adopts required and
forbidden retrieval assertions for project isolation, supersession,
abstention, temporal state, and latency, with PrecisionMemBench as its
primary design reference. The project retains a
different architecture combining SQLite, full-text retrieval, vector
retrieval, and reciprocal-rank fusion. This indicates that the evaluation
contract can inform testing across architectures without requiring
adoption of Tenure's retrieval implementation.

These uses provide evidence
that retrieval precision is an operational system property not fully
captured by recall-oriented or answer-level evaluation.

	\section{Conclusion}
	\label{sec:conclusion}
	
	Persistent LLM memory is a state-management problem as well as a search
	problem. Across 13 configurations, comparison systems often preserved
	recall by returning broad candidate sets but failed to isolate the
	required belief. Tenure instead resolves ownership, scope, lifecycle,
	and identity before inference, passing all 89 benchmark cases.
	
	By scoring retrieved belief IDs before answer generation,
	PrecisionMemBench exposes scope leakage, supersession leakage, and
	topic-drift contamination that answer-quality evaluation can conceal.
	The benchmark has also supported provider reruns, implementation changes,
	and independent regression tracking. Its results should be interpreted
	as reproducible measurements of specific versions and configurations,
	not permanent judgments about the systems evaluated.

\end{document}